\documentclass[prd,reprint,twocolumn,nofootinbib,showpacs,preprintnumbers,superscriptaddress,amsmath,amssymb]{revtex4}
\usepackage{amsmath}
\usepackage{lmodern}
\usepackage{graphicx}

\def\MeV{\text{ MeV}}

\def\beq{\begin{equation}} 
\def\enq{\end{equation}} 
\def\beqa{\begin{eqnarray}} 
\def\enqa{\end{eqnarray}} 

\begin{document}

\title{\bf \boldmath $D_{s0}^{\ast\pm}(2317)$ and $KD$ scattering from $B^0_s$ decay}

\author{Miguel~Albaladejo}
\email{Miguel.Albaladejo@ific.uv.es}
\affiliation{ Departamento de F\'{\i}sica Te\'orica and IFIC, Centro
  Mixto Universidad de Valencia-CSIC, Institutos de Investigaci\'on de
  Paterna, Aptdo. 22085, 46071 Valencia, Spain }

\author{Marina~Nielsen}
\email{mnielsen@if.usp.br}
\affiliation{Instituto de F\'{\i}sica, Universidade de S\~{a}o Paulo,
  C.P. 66318, 05389-970, S\~{a}o Paulo, SP, Brazil}

\author{Eulogio~Oset}
\email{oset@ific.uv.es}
\affiliation{ Departamento de F\'{\i}sica Te\'orica and IFIC, Centro
  Mixto Universidad de Valencia-CSIC, Institutos de Investigaci\'on de
  Paterna, Aptdo. 22085, 46071 Valencia, Spain }

\begin{abstract}
We study the $\bar{B}_s^0\to D_s^-(DK)^+$ weak decay, and  look at the $DK$ invariant mass 
distribution with the aim of obtaining relevant information on the nature of the 
$D_{s0}^{\ast+}(2317)$ resonance. We make a simulation of the experiment using the actual mass 
of the $D_{s0}^{\ast+}(2317)$ resonance and recent lattice QCD relevant parameters of the $KD$ 
scattering amplitude.  We then solve the inverse problem of obtaining the $KD$ amplitude from 
these synthetic data, to which we have added a 5\% or 10\% error. We prove that one can obtain 
from these "data" the existence of a bound $KD$ state, the $KD$ scattering length and effective 
range, and most importantly, the $KD$ probability in the wave function of the bound state 
obtained, which was found to be largely dominant from the lattice QCD results. This means that 
one can obtain information on the nature of the $D_{s0}^{\ast+}(2317)$ resonance from the 
implementation of this experiment, in the line of finding the  structure of resonances, which 
is one of the main aims in hadron spectroscopy. 
\end{abstract}

\maketitle

\section{Introduction}

The very narrow charmed scalar meson $D_{s0}^{\ast+}(2317)$ was first observed in 
the $D_s^+\pi^0$ channel by the BABAR Collaboration \cite{babar} and its
existence was confirmed by CLEO \cite{cleo}, BELLE \cite{belle1} and
FOCUS \cite{focus} Collaborations. Its mass was commonly measured as
$2317 \MeV$, which is approximately $160 \MeV$ below the prediction of 
the very successful quark model for the charmed mesons \cite{god}. 
Due to its low mass, the structure of the meson $D_{s0}^{\ast\pm}(2317)$ has been 
extensively debated. It has been interpreted as 
a $c\bar{s}$ state \cite{dhlz,bali,ukqcd,ht,nari}, two-meson molecular 
state \cite{bcl,szc,Kolomeitsev:2003ac,Guo:2006fu,gamermann,Guo:2009ct,Cleven:2010aw,Cleven:2014oka}, $K-D$- mixing \cite{br},  
four-quark states \cite{ch,tera,mppr,nos} or a mixture between two-meson
and four-quark states \cite{bpp}. Additional support to the molecular 
interpretation  came recently   from lattice QCD simulations \cite{sasa1,Liu:2012zya,sasa2,sasa3}. 
In previous lattice studies of the $D_{s0}^{\ast}(2317)$, it was treated as a 
conventional quark-antiquark state and  no states with 
the correct mass  (below  the  $KD$ threshold) were found. In Refs.~\cite{sasa1,sasa2}, with the introduction of $KD$  meson operators and using the 
effective range formula, a bound state is obtained about 40 MeV below the $KD$
threshold. The fact that the bound state appears with the $KD$ interpolator may be interpreted
as a possible $KD$ molecular structure, but more precise statements cannot be
done. In Ref.~\cite{Liu:2012zya} lattice QCD results for the $KD$ scattering length are extrapolated to physical pion masses by means of unitarized chiral perturbation theory, and by means of the Weinberg compositeness condition \cite{Weinberg:1965zz,Baru:2003qq} the amount of $KD$ content in the $D_{s0}^*(2317)$ is determined, resulting in a sizable fraction of the order of 70\% within errors. A reanalyis of the lattice spectra of Refs.~\cite{sasa1,sasa2} has been recently done in 
Ref.~\cite{sasa3}, going beyond the effective range approximation and making use of the three 
levels of Refs.~\cite{sasa1,sasa2}. As a consequence, one can be more quantitative about the nature of the $D_{s0}(2317)$, which appears with a $KD$ component of about 70\%, within errors. 

In addition to these lattice results, and more precise ones that should be available in the 
future, it is very important to have some experimental data that could be used to test the 
internal structure of this exotic state.

Here we propose to use the experimental $KD$ invariant mass distribution of the weak decay of 
$\bar{B}_s^0\to D_s^-(DK)^+$\footnote{Throughout this work, the notation $(DK)^+$ refers to the 
isoscalar combination $D^0 K^+ + D^+ K^0$.} in order to obtain information about the internal 
structure of the  $D_{s0}^{\ast+}(2317)$ state.
There are not yet experimental data for the decay $\bar{B}^0_s \to D_s^-(DK)^+$.
However, since the branching fractions  for the decays $\bar{B}^0_s \to D_s^{*+}D_s^{*-}$ and
$\bar{B}^0_s \to D_s^{+}D_s^{*-}+D_s^{*+}D_s^{-}$ are respectively 1.85\% and
1.28\%, we believe that the branching fraction for the $\bar{B}^0_s \to D_s^-D_{s0}^{\ast+}$
decay, should not be so different from that and  it will be seen through the
channel $\bar{B}^0_s \to D_s^-(DK)^+$. This is why it is really
important to have theoretical predictions for the $DK$ invariant mass
distribution that considers the formation of the $D_{s0}^{\ast+}(2317)$ state. At this point,
 it is worth stressing that recently, in the reactions $B^0 \to D^- D^0 K^+$ and $B^+ \to 
\bar{D}^0 D^0 K^+$ studied by the BABAR Collaboration \cite{Lees:2014abp}, an enhancement in 
the invariant $DK$ mass in the range $2.35-2.50\ \text{GeV}$ is observed, which could be 
associated with this $D_{s0}^{\ast+}(2317)$ state. It is also interesting to quote that in a different reaction, $B^0_s \to \bar D^0 K^- \pi^+$,  the LHCb Collaboration also finds an enhancement close to the $KD$ threshold in the $\bar D^0 K^-$ invariant mass distribution, which is partly associated to the $D_{s0}^*(2317)$ resonance \cite{gershot}.

\begin{figure*}\centering
\includegraphics[height=4.5cm,keepaspectratio]{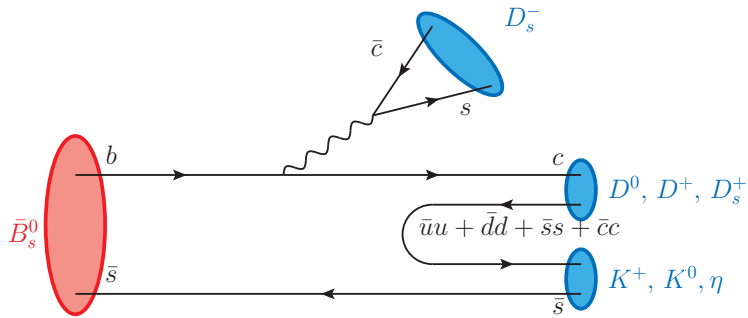}
\caption{Mechanism for the decay $\bar{B}^0_s \to D_s^-(DK)^+$.\label{fig:diag}}
\end{figure*}

In Fig.~\ref{fig:diag} we show the mechanism for the decay $\bar{B}_s^0\to D_s^-
(DK)^+$. The idea is to take the dominant mechanism for the weak decay of the
$\bar{B}_s^0$ into $D_s^-$ plus a primary $c\bar{s}$ pair. The hadronization of 
the initial $c\bar{s}$ pair is achieved by inserting a $q\bar q$ pair with the 
quantum numbers of the vacuum: $u\bar u +d\bar d +s\bar s + c \bar c$, as shown in 
Fig.~\ref{fig:diag}. Therefore, the $c\bar{s}$ pair is hadronized into a pair of 
pseudoscalar mesons. This pair of pseudoscalar mesons is then allowed to interact
to produce the $D_{s0}^{\ast+}(2317)$ resonance, which is considered here as 
mainly a $DK$ molecule \cite{gamermann}. The idea is similar to the one used in 
Ref.~\cite{liang} for the formation of the $f_0(980)$ and $f_0(500)$ scalar resonances 
in the decays of $B^0$ and $B^0_s$.

The paper is organized as follows. In Section~\ref{sec:formalism} we settle the formalism for 
our study. Namely, in Subsection~\ref{subsec:Ampli} we study the $(DK)^+$ elastic scattering 
amplitude, and in Subsection~\ref{subsec:DiffWid} we study the differential decay width for the 
process $B^0_s \to D_s^- (DK)^+$. As said before, there is not yet experimental information 
concerning the differential decay width for this process. For this reason, we will have to 
generate synthetic data for this decay in order to explore if this reaction is suitable for the 
study of the $(DK)^+$ final state interactions and the $D_{s0}^{\ast+}(2317)$ bound state. The 
generation and analysis of these synthetic data, which constitutes the results of the work, are 
done in Subsection~\ref{sec:results}. Conclusions are delivered in Section~\ref{sec:conclusions}.

\section{Formalism}\label{sec:formalism}

In this work the influence of the presence of the $D_{s0}^{\ast+}(2317)$ in the process 
$\bar{B}^0_s \to D_s^-(DK)^+ $ is investigated. The $D_{s0}^{\ast+}(2317)$ is considered mainly 
as a bound state of the $DK$ system, so we address the elastic $DK$ scattering amplitude in 
Subsection~\ref{subsec:Ampli}. Then, the differential decay width for the $\bar{B}^0_s \to 
D_s^-(DK)^+$ reaction in terms of the $DK$ invariant mass is considered in 
Subsection~\ref{subsec:DiffWid}.
\subsection{\boldmath Elastic $DK$ scattering amplitude}\label{subsec:Ampli}
Let us start by discussing the $S$-wave amplitude for the isospin $I=0$ $DK$ elastic scattering, 
which we denote $T$. It can be written as
\begin{equation}\label{eq:TEFT}
T^{-1}(s) = V^{-1}(s)-G(s)\Rightarrow T(s)=V(s)(1+G(s)T(s))~,
\end{equation}
where $G(s)$ is a loop function bearing the unitary or right hand cut,
\begin{equation}\label{eq:GloopIntegral}
G(s) \equiv 
i \int \frac{d ^{4} l}{(2 \pi)^{4}} 
\frac{1}{l^{2} - m_{K}^{2} + i \epsilon } 
\frac{1}{(q - l)^{2} - m_{D}^2 + i \epsilon} ,
\end{equation}
and $s=q^2$ is the invariant mass squared of the $DK$ system. This function needs to be 
regularized, and this is accomplished in this work by means of a subtraction constant, $a(\mu)$. 
In this way, the $G$ function can be written as:
\begin{align}
& 16\pi^2 G(s) = a(\mu) +  \log\frac{m_D m_K}{\mu^2} + \frac{\Delta}{2s}\log\frac{m_D^2}{m_K^2} 
\nonumber\\
&+ \frac{\nu}{2s} 
\left( 
\log\frac{s-\Delta+\nu}{-s+\Delta+\nu} + 
\log\frac{s+\Delta+\nu}{-s-\Delta+\nu}
\right) \label{eq:GloopSubtracted}~, \\
& \Delta = m_D^2-m_K^2~, \quad \nu = \lambda^{1/2}(s,m_D^2,m_K^2)~,\nonumber
\end{align}
where $\lambda(x,y,z) = x^2+y^2+z^2-2xy-2yz-2zx$ is the K\"allen or triangle function. In 
Eq.~\eqref{eq:TEFT}, $V(s)$ is the potential, typically extracted from some effective field 
theory, although a different approach will be followed here (see below).

The amplitude $T(s)$ can also be written in terms of the phase shift $\delta(s)$ and/or 
effective range expansion parameters,
\begin{equation}
T(s)= -\frac{8\pi\sqrt{s}}{p_K \text{ctg}\delta - i p_K}\simeq-\frac{8\pi \sqrt{s}}{\displaystyle 
\frac{1}{a}+\frac{1}{2}r_0 p_K^2 -i p_K} ,
\label{ampl}
\end{equation}
with
\begin{equation}\label{eq:pk}
p_K(s) = \frac{\lambda^{1/2}(s,M_K^2,M_D^2)}{2\sqrt{s}}~,
\end{equation}
the momentum of the $K$ meson in the $DK$ center of mass system. Above, $a$ and $r_0$ are the 
scattering length and the effective range, respectively. 

In this channel and linked to it we find the $D_{s0}^{\ast+}(2317)$ resonance, the object of 
study of this paper, below the $DK$ threshold, the latter being located roughly above 
$2360\ \text{MeV}$. This means that the amplitude has a pole at the squared mass of this state, 
$M^2 \equiv s_0$, so that, around the pole,
\begin{equation}\label{eq:Tpole}
T(s) = \frac{g^2}{s - s_0} + \text{regular terms},
\end{equation}
being $g$ the coupling of the state to the $DK$ channel. From Eqs.~\eqref{eq:TEFT} and 
\eqref{eq:Tpole}, we see that (the following derivatives are meant to be calculated at $s=s_0$):
\begin{equation}
\frac{1}{g^2} = \frac{\partial T^{-1}(s) }{\partial s} = \frac{\partial V^{-1}(s)}{\partial s} 
- \frac{\partial G(s)}{\partial s}~.
\end{equation} 
We have thus the following exact sum rule,
\begin{equation}\label{eq:sumrule}
1 = g^2\left(- \frac{\partial G(s)}{\partial s} + \frac{\partial V^{-1}(s)}{\partial s} \right)~.
\end{equation}
In Ref.~\cite{Gamermann:2009uq} it has been shown, as a generalization of the Weinberg 
compositeness condition \cite{Weinberg:1965zz} (see also Ref.~\cite{Sekihara:2014kya} and 
references therein), that the probability $P$ of finding the channel under study (in this case, 
$DK$) in the wave function of the bound state is given by:
\begin{equation}\label{eq:P_def}
P = - g^2 \frac{\partial G(s)}{\partial s}~,
\end{equation}
while the rest of the r.h.s. of Eq.~\eqref{eq:sumrule} represents the probability of other 
channels, and hence the probabilities add up to 1. Finally, if one has an energy independent 
potential the second term of Eq.~\eqref{eq:sumrule} vanishes, and then $P=1$, that is, the bound 
state is purely given by the channel under consideration. In Ref.~\cite{Gamermann:2009uq}, these 
ideas are generalized to the coupled channels case.

Let us now apply these ideas to the case of $DK$ scattering. From Eq.~\eqref{eq:TEFT} it can be 
seen that the existence of a pole implies
\begin{align}
V^{-1}(s) & \simeq G(s_0) + \alpha (s-s_0) + \cdots~,\\
\alpha & \equiv \left. \frac{\partial V^{-1}(s)}{\partial s} \right\rvert_{s=s_0}~,
\end{align}
in the neighbourhood of the pole. If we assume that the energy dependence is smooth enough to 
allow us to retain only the linear term in $s$, we can now write the amplitude as
\begin{equation}\label{eq:expansionTG}
T^{-1}(s) = G(s_0) - G(s) + \alpha(s-s_0)~,
\end{equation}
and the sum rule in Eq.~\eqref{eq:sumrule} becomes:
\begin{equation}\label{eq:sumrule2}
P_{DK} = 1 - \alpha g^2~.
\end{equation}
In this way, the quantity $\alpha g^2$ represents the probability of finding other components 
beyond $DK$ in the wave function of $D_{s0}^{\ast+}(2317)$. The following relation can also be 
deduced from Eqs.~\eqref{eq:sumrule2} and \eqref{eq:P_def}:
\begin{equation}\label{eq:rel_P_alpha}
\alpha = - \frac{1-P_{DK}}{P_{DK}} \left. \frac{\partial G(s)}{\partial s} \right\rvert_{s=s_0}.
\end{equation}

We can now link this formalism with the results of Ref.~\cite{sasa3}, where a reanalysis is done 
of the energy levels found in the lattice simulations of Ref.~\cite{sasa2}. In Ref.~\cite{sasa3},
 the following values for the effective range parameters are found:
\begin{equation}\label{eq:a0_r0_valuesSasa}
a_0 = -1.4 \pm 0.6\ \text{fm}~,\quad r_0 = -0.1 \pm 0.2\ \text{fm}~.
\end{equation}
Also, in studying the $D_{s0}^{\ast+}(2317)$ bound state, a binding energy $B = M_D + M_K - 
M_{D_{s0}^{\ast+}} = 31 \pm 17\ \text{MeV}$ is found in Ref.~\cite{sasa3}. The probability 
$P_{DK}$ is also studied, and the value $P_{DK} = 0.72 \pm 0.12$ is found. Hence, for our 
analysis, in which synthetic data for the reaction $\bar{B}^0_s \to (DK)^+ D_s^-$ will be 
generated, we can start from the hypothesis that a bound state exists in the $DK$ channel, with 
a mass $M_{D_{s0}^{\ast+}} = 2317\ \text{MeV}$ (the nominal one), and with a probability 
$P_{DK}=0.75$. This implies, from Eq.~\eqref{eq:rel_P_alpha}, the value $\alpha = 
2.06\cdot 10^{-3}\ \text{GeV}^{-2}$. Finally, for the subtraction constant in the $G$ function, 
Eq.~\eqref{eq:GloopSubtracted}, we shall take, as in Ref.~\cite{gamermann}, the value 
$a(\mu) = -1.3$ for $\mu = 1.5\ \text{GeV}$. Note that $\partial G(s)/\partial s$ does not depend on $\mu$ or $a(\mu)$.

\subsection{\boldmath Decay amplitude and invariant $DK$ mass distribution in the $\bar{B}^0_s \to 
D_s^-(DK)^+$ decay}\label{subsec:DiffWid}

We now discuss the amplitude for the decay $B^0_s \to D_s^-(DK)^+$ decay, and its relation to 
the $DK$ elastic scattering amplitude studied above. The basic mechanism for this process is 
depicted in Fig.~\ref{fig:diag}, where, from the $\bar{s}b$ initial pair constituting the 
$B^0_s$, a $\bar{c}s$ pair and a $\bar{s}c$ pair are created. The first pair produces the 
$D_s^-$, and the $DK$ state arises from the hadronization of the second pair. Let us consider 
in some more detail the hadronization mechanism. To construct a two meson final state, the 
$c\bar{s}$ pair has to combine with another $\bar{q}q$ pair created from the vacuum. We 
introduce the following matrix,
\begin{align}
M & = v \bar{v} = 
\left(\begin{array}{c}
u \\ d \\ s \\ c
\end{array}\right)
\left(\begin{array}{cccc}
\bar{u} & \bar{d} & \bar{s} & \bar{c}
\end{array}\right) \nonumber\\
& = \left( \begin{array}{cccc} 
u\bar{u} & u\bar{d} & u\bar{s} & u\bar{c} \\
d\bar{u} & d\bar{d} & d\bar{s} & d\bar{c} \\
s\bar{u} & s\bar{d} & s\bar{s} & s\bar{c} \\
c\bar{u} & c\bar{d} & c\bar{s} & c\bar{c} \\
\end{array}\right)~,
\end{align}
which fulfils:
\begin{align}
M^2 & = (v \bar{v})(v \bar{v}) = v (\bar{v} v) \bar{v} \nonumber\\ 
& = \left( \bar{u}u + \bar{d}d + \bar{s}s + \bar{c}c \right) M~.
\end{align}
The first factor in the last equality represents the $\bar{q}q$ creation. This matrix $M$ is in 
correspondence with the meson matrix $\phi$:
\begin{equation}
\phi = \left( \begin{array}{cccc} 
\frac{\eta}{\sqrt{3}} + \frac{\pi^0}{\sqrt{2}} + \frac{\eta'}{\sqrt{6}} & \pi^+ & K^+ & \bar{D}^0 \\
\pi^- & \frac{\eta}{\sqrt{3}} - \frac{\pi^0}{\sqrt{2}} + \frac{\eta'}{\sqrt{6}} & K^0 & D^- \\
K^- & \bar{K}^0 & \frac{\sqrt{2}\eta'}{\sqrt{3}}-\frac{\eta}{\sqrt{3}} & D_s^- \\
D^0 & D^+ & D_s^+ & \eta_c
\end{array}\right)~.
\end{equation}
\begin{figure*}\centering
\includegraphics[height=3.5cm,keepaspectratio]{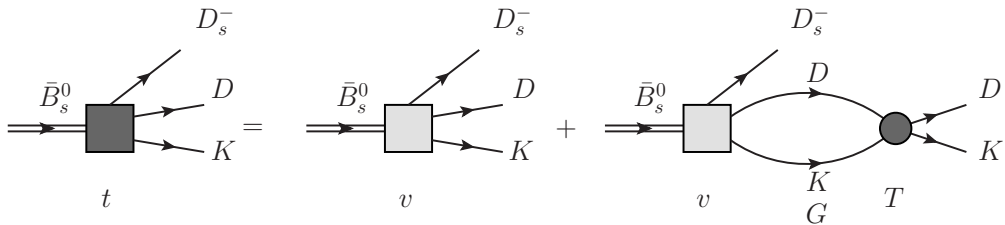}
\caption{Diagrammatical interpretation of Eq.~\eqref{eq:DecayAndFSI}, in which $DK$ final state 
interaction is taken into account for the decay $\bar{B}^0_s \to D_s^-(DK)^+$. The dark square 
represents the amplitude $t$ for the process, in which the final state interaction is already 
taken into account. The light square represents the bare vertex for the process, denoted by $v$. 
Finally, the circle represents the hadronic amplitude for the elastic $DK$ 
scattering.\label{fig:diag2}}
\end{figure*}
The hadronization of the $c\bar{s}$ pair proceeds then through the matrix element 
$\left(M^2\right)_{43}$, which in terms of mesons reads:
\begin{equation}
(\phi^2)_{43} = K^+ D^0 + K^0 D^+ + \cdots~,
\end{equation}
where only terms containing a $KD$ pair are retained, since coupled channels are not considered 
in this work. We note that this $KD$ combination has $I=0$, as it should, since it is produced 
from a $c\bar{s}$, which has $I=0$, and the strong interaction hadronization which conserves 
isospin (the $\bar{q} q$ with the quantum numbers of the vacuum has $I=0$).

Let $t$ be the full amplitude for the process $B^0_s \to D_s^- (DK)^+$, which already takes into 
account the final state interaction of the $DK$ pair. Also, let us denote by $v$ the bare vertex
 for the same reaction. To relate $t$ and $v$, that is, to take into account the final state 
interaction of the $DK$ pair, as sketched in Fig.~\ref{fig:diag2}, we write: 
\begin{equation}\label{eq:DecayAndFSI}
t = v + v G(s) T(s)=v(1+G(s)T(s))~.
\end{equation}
From Eq.~\eqref{eq:TEFT}, the previous equation can also be written as:
\begin{equation}
t = v \frac{T(s)}{V(s)}~.
\end{equation}
Because of the presence of the bound state below threshold, this process will depend strongly on 
$s$ in the kinematical window ranging from threshold to $100\ \text{MeV}$ above it, so we can 
safely assume that $t$ depends only on $s$. Hence, the differential width for the process under 
consideration is given by:
\begin{equation}\label{eq:FinalDifWid}
\frac{\text{d} \Gamma}{d\sqrt{s}} = \frac{1}{32\pi^2 M_{\bar{B}^0_s}^2} p_{D_s^-} p_K \left\lvert
 t \right\rvert^2 = \mathcal{C} p_{D_s^-} p_K \left\lvert \frac{T(s)}{V(s)} \right\rvert^2~,
\end{equation}
where the bare vertex $v$ has been absorbed in $\mathcal{C}$, a global (but otherwise not 
relevant) constant, and where $p_K$ is given in Eq.~\eqref{eq:pk} and $p_{D^-_s}$ is the 
momentum of the $D_s^-$ meson in the rest frame of the decaying $\bar{B}^0_s$, given by:
\begin{equation}
p_{D^-_s} = \frac{\lambda^{1/2}(M_{\bar{B}^0_s}^2,M_{D^-_s}^2,s)}{2 M_{\bar{B}^0_s}}~.
\end{equation}

\section{Results}\label{sec:results}

\begin{table}[t]\centering
\renewcommand{\arraystretch}{1.3}
\begin{tabular}{|c|cll|} \hline
 & Central Value & $5\ \%$ & $10\ \%$ \\ \hline
$10^3\ \alpha\ (\text{GeV}^{-2})$ & $2.06$ & $^{+0.17}_{-0.40}$ & $^{+0.10}_{-1.09}$
 \\ \hline
$M_{D^\ast_{s0}}\ (\text{MeV})$ & $2317$ & $^{+14}_{-24}$ & $^{+21}_{-73}$ \\ \hline
$a(\mu)$ & $-1.30$ & $^{+0.15}_{-0.37}$ & $^{+0.27}_{-0.49}$ \\ \hline
$|g|\ (\textrm{GeV})$ & $11.0$ & $^{+1.0}_{-0.6}$ & $^{+2.2}_{-1.1}$ \\ \hline
$a_0\ (\textrm{fm})$ & $-1.0$ & $^{+0.2}_{-0.2}$ & $^{+0.4}_{-0.5}$ \\ \hline
$r_0\ (\textrm{fm})$ & $-0.14$ & $^{+0.06}_{-0.03}$ & $^{+0.16}_{-0.04}$ \\ \hline
$P_{DK}$ & $0.75$ & $^{+0.07}_{-0.06}$ & $^{+0.16}_{-0.11}$ \\ \hline
\end{tabular}
\caption{Fitted parameters ($\alpha$, $M_{D_{s0}^{\ast+}}$ and $a(\mu)$)  and 
predicted quantities ($|g|$, $a_0$, $r_0$, $P_{DK}$) for $\mu=1.5\ \text{GeV}$. The second column shows the 
central value of the fit, whereas the third (fourth) column presents the errors 
(estimated by means of MC simulation) when the experimental error is 5\% (10\%). 
\label{tab:res}}
\end{table}

We want to investigate the presence of the $D_{s0}^{\ast+}(2317)$ state in the $(DK)^+$ 
scattering amplitude. In order to explore the sensitivity of the decay $\bar{B}^0_s \to  (DK)^+ 
D^-_s$ to the presence of this bound state, we generate synthetic data from our theory for the 
differential decay width for the process with Eqs.~\eqref{eq:FinalDifWid} and 
\eqref{eq:expansionTG}. We generate 10 synthetic points in a range of $100\ \text{MeV}$ starting 
from threshold. To each centroid, we assign the value obtained with the central values explained 
in Subsection~\ref{subsec:Ampli} ($10^3\alpha = 2.06\ \text{GeV}^{-2}$, $a(\mu)=-1.3$, and 
$M_{D_{s0}^{\ast+}} = 2317\ \text{MeV}$). We shall study two different cases, in which each 
experimental point is given an error of a 5\% or a 10\% of the highest value of the differential 
decay width. Taking these synthetic data as experiment-given data, we perform the inverse 
problem of analysing them with our theory.  Obviously, the reproduction of the data must be 
perfect, but we recall that the scope here is to investigate the experimental accuracy that is 
actually needed to obtain reliable values for the quantities fitted or predicted from our theory 
($M_{D_{s0}^{\ast+}}$, $a_0$, $r_0$, and $P_{DK}$). The analysis of these synthetic data goes as 
follows. We generate around $2\cdot 10^3$ sets of  random experimental points, in which each 
centroid is varied around its theoretical value according to a Gaussian distribution with the 
error given to each point. For each of these sets of random points, the parameters are fitted to 
the data. After the whole run, a central range, containing a 68\% of the values of the considered
 quantities (the differential decay width, the fitted parameters, and the predicted values) is 
retained. It is worth stressing here that, since the centroid of the experimental point in each 
set of random experimental points is varied, a good reproduction of the random synthetic data is 
quite natural but not completely trivial. 

\begin{figure}[t]\centering
\includegraphics[height=5cm,keepaspectratio]{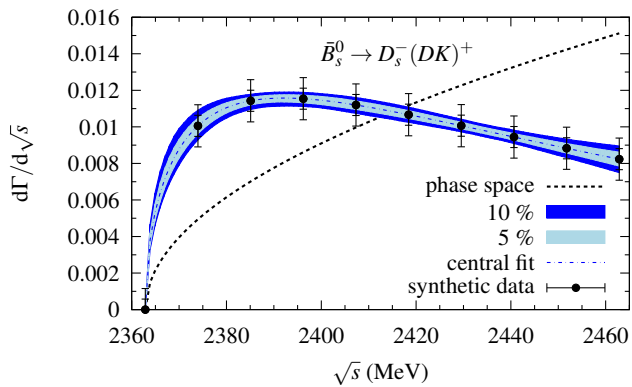}
\caption{Differential decay width for the reaction $\bar{B}^0_s \to D^-_s (DK)^+$. The synthetic 
data (generated as explained in the text) are shown with black points. The smaller (larger) 
error bars correspond to a 5\% (10\%) experimental error. The dash-dotted line represents the 
theoretical prediction obtained with the central values of the fit. The light (dark) bands 
correspond to the estimation of the error (by means of a MC simulation) when fitting the data 
with 5\% (10\%) experimental error. The dashed line corresponds to a phase space distribution
normalized to the same area in the range examined.\label{fig:DifWid}}
\end{figure}

The generated synthetic data are shown in Fig.~\ref{fig:DifWid}. As explained, they have two 
different error bars, the smaller one corresponding to a 5\% experimental error and the larger 
one to a 10\%. As commented 
above, they exactly match the central curve (dash-dotted line) produced with the central 
parameters of the theory. A solely phase 
space distribution ({\it i.e.}, a differential decay width proportional to $p_{D_s^-}p_K$, but 
with no other kinematical dependence of dynamical origin) is also shown in the figure (dashed 
line). The first important information to be extracted from the figure is that the data are 
clearly incompatible with this phase space distribution. This points to the presence of a 
resonant or bound state or, at least, to some strong final state interactions. Two error bands 
are shown in the same figure, the lighter and smaller (darker and larger) one corresponding to a 
5\% (10\%) experimental error.  The fitted parameters ($a(\mu)$, $M_{D_{s0}^{\ast+}}$, and 
$\alpha$) are shown in Table~\ref{tab:res}, together with their errors.\footnote{To avoid 
unphysical values of the fitted parameters $a(\mu)$ and $\alpha$, which could numerically 
reproduce each set of the randomly generated experimental points, they are restricted to vary 
within a sensible range, but making sure that this range is larger than the error obtained for 
these two parameters and shown in Table~\ref{tab:res}.} Note that, with a 5\% experimental error,
 we get $M_{D_{s0}^{\ast+}} = 2317^{+14}_{-24}\ \text{MeV}$, and if the error is increased to 
10\%, the value is $M_{D_{s0}^{\ast+}} = 2317^{+21}_{-73}\ \text{MeV}$. Here we are mainly 
concerned with the upper error, in the sense that it is the one that defines if the bound state 
is clearly below the $DK$ threshold (which is slightly above $2360\ \text{MeV}$) or not. 
Considering this error, we see that the mass obtained is well below the threshold, at the level 
of $2\sigma$ ($3\sigma$) for the case of a 5\% (10\%) experimental error. This is a good 
information: experimental data with a 10\% error, which is clearly feasible with nowadays 
experimental facilities, can clearly determine the presence of this below threshold state 
$D_{s0}^{\ast+}(2317)$.

We can also determine $P_{DK}$, the probability of finding the $DK$ channel in the 
$D_{s0}^{\ast+}(2317)$ wave function. It is shown in the last row of Table~\ref{tab:res}. As 
stated, the central value $P_{DK}=0.75$ is the same as the initial one, but we are here 
interested in the errors, which are small enough even in the case of a 10\% experimental error. 
This means that with the analysis of such an experiment one could address with enough accuracy 
the question of the molecular nature of the state ($D_{s0}^{\ast+}(2317)$, in this case).

Finally, it is also possible to determine other parameters related with $DK$ scattering, such as 
the scattering length ($a_0$) and the effective range ($r_0$). They are also shown in 
Table~\ref{tab:res}. They are compatible with the lattice QCD studies presented in 
Refs.~\cite{sasa2, sasa3}. Namely, the results from Ref.~\cite{sasa3} are shown in 
Eqs.~\eqref{eq:a0_r0_valuesSasa}, and their mutual compatibility is clear.

\section{Conclusions}\label{sec:conclusions}

In the present work we have selected a reaction which is both Cabbibo and color favored, the 
$\bar{B}_s^0\to D_s^-(DK)^+$ weak decay, and have looked at the $DK$ invariant mass distribution 
from where we expect to obtain relevant information on the nature of the $D_{s0}^{\ast+}(2317)$ 
resonance when actual data are available. For this purpose we have performed a simulation of the 
experiment taking information from experiment about the mass of the $D_{s0}^{\ast+}(2317)$ 
resonance and from a recent QCD lattice analysis on an analytical representation of the $KD$ 
scattering amplitude.  This information has served us to make predictions on the shape of the 
$KD$ invariant mass distribution close to the $KD$ threshold.  After that we have taken these 
results and we have assumed they are actual "experimental data", associating to them an 
"experimental error" of 5\% or 10\%. Then we have made a fit to these "synthetic data" in order 
to extract from there the  $KD$ scattering amplitude, above and below threshold. We prove  that 
with both errors, typical of present experimental data of spectra in $B$ decays, one can obtain 
the $KD$ scattering amplitude with enough precision to predict that there is a $KD$ bound state. 
We also predict the scattering lenght and effective range of the $KD$ interaction and, very 
important, we show that we can predict, with relatively small error, the probability of the 
mesonic $KD$ component in the wave function of the $D_{s0}^{\ast+}(2317)$ resonance. From the QCD 
lattice results one induces about 70\% probability and we show that this number can be obtained 
from the analysis of the $B$ decay spectra with sufficient precission to make the number 
significative of the main nature of the $D_{s0}^{\ast+}(2317)$ resonance as a basically $KD$ 
molecular state with a smaller mixture of other components. 

  The study done here should stimulate the implementation of the experiment, for which we have 
made estimates of a relatively large branching fraction. 

\acknowledgements
This work is partly supported
by the Spanish Ministerio de Economia y Competitividad and European
FEDER funds under the contract number FIS2011-28853-C02-01 and
FIS2011-28853-C02-02, and the Generalitat Valenciana in the program
Prometeo II-2014/068. We acknowledge the support of the European
Community-Research Infrastructure Integrating Activity Study of
Strongly Interacting Matter (acronym HadronPhysics3, Grant Agreement
n. 283286) under the Seventh Framework Program of the EU. 
M.~A. acknowledges financial support from the ``Juan de la Cierva'' program 
(reference 27-13-463B-731) from the Spanish Government through the Ministerio de 
Econom\'ia y Competitividad. M.N. acknowledges the IFIC for the hospitality and support during 
her stay in Valencia and also support from CNPq and FAPESP.

\newpage

\bibliographystyle{plain}
  
  \end{document}